\documentclass{astlb}
\usepackage{graphicx} 
\usepackage{amsmath} 
\usepackage{epsfig,graphics} 
\usepackage{rotating} 
\usepackage{epstopdf}
\usepackage{pdfpages}
\usepackage{longtable}
\usepackage{caption}
\usepackage{dcolumn}
\usepackage{pdflscape}
\usepackage{comment}
\usepackage{color}
\usepackage{tcolorbox}
\usepackage{xspace}
\usepackage{multirow}
\usepackage{hyperref}

\begin{document} 

\journalinfo{2024}{50}{11}{676}[686]
\title{An X-ray and Optical Study of the Dwarf Nova Candidate OGLE-BLG-DN-0064}

\author{
A. B. Sibgatullin\address{1}\email{absibgatullin@kpfu.ru}, 
V. I. Dodon\address{1}, 
I. I. Galiullin\address{1}, 
A. I. Kolbin\address{1,2}, 
V. V. Shimansky\address{2}, 
A. S. Vinokurov\address{2}
\addresstext{1}{Kazan Federal University, Kazan, 420000 Russia}
\addresstext{2}{Special Astrophysical Observatory, Russian Academy of Sciences, Nizhnii Arkhyz, 369167 Russia}}
 
\shortauthor{Sibgatullin et al.}  
\shorttitle{X-ray and Optical Study of OGLE-BLG-DN-0064} 
\submitted{November 13, 2024; revised December 5, 2024; accepted December 12, 2024}

\begin{abstract}  
The source OGLE-BLG-DN-0064 (hereafter OGLE64) was classified as a potential dwarf nova based on its regular outburst activity revealed by the OGLE optical survey. In this paper, we investigate the X-ray and optical emissions from the source OGLE64 based on archival \textit{Chandra} and \textit{Swift} X-ray data and our optical observations with the 6-m BTA telescope at the Special Astrophysical Observatory of the Russian Academy of Sciences. OGLE64 shows an X-ray luminosity \( L_X \approx 1.6 \times 10^{32} \, \text{erg s}^{-1} \) and a high X-ray-to-optical flux ratio \( F_X / F_{\text{opt}} \approx 1.5 \), typical for accreting white dwarfs. The X-ray spectrum of OGLE64 is better fitted by the models of a power law with a photon index \(\Gamma \approx 1.9\) and an optically thin plasma with a temperature \( kT \approx 6.4 \, \text{keV} \). The optical spectrum shows hydrogen and neutral helium emission lines, in some of which a double-peaked structure is observed. An analysis of the outburst activity of OGLE64, based on data from the OGLE, ZTF, ATLAS, and ASAS-SN optical surveys, has revealed superoutbursts with a characteristic supercycle \( P_{\text{super}} \approx 400 \, \text{days} \). We found no significant variability in either the X-ray or optical light curves of OGLE64 that could be associated with the change in the visibility conditions for the emitting regions at different orbital phases. Our estimates of the orbital period of the system by indirect methods show that the period probably lies in the range \( P_{\text{orb}} \sim 1.5 - 3.5 \, \text{h} \). The properties of the X-ray and optical emissions from OGLE64 lead us to conclude that the system is an SU UMa-type dwarf nova.

\textbf{DOI:} 10.1134/S1063773725700033

\end{abstract}


\section{Introduction}

Cataclysmic variables (CVs) are close binary systems that consist of an accreting white dwarf (WD) and a donor star filling its Roche lobe \citep{1995cvs..book.....W}. CVs are X-ray sources with typical luminosities \( L_X \sim 10^{30} - 10^{33} \, \text{erg s}^{-1} \) \citep{2017PASP..129f2001M}, making it possible to search for and investigate such systems with space X-ray observatories (see, e.g., \citealt{1984MNRAS.206..879C, 1995A&A...297L..37H, 2020NewAR..9101547L}).

Nonmagnetic CVs exhibiting regular outburst activity are called dwarf novae. Dwarf novae are separated by the shape of their outbursts into several subclasses. The so-called normal outbursts, which have a duration of several days and an amplitude of \( 2^m - 5^m \), are observed in U Gem-type stars \citep{1975IAUS...67..357G}. Z Cam-type systems exhibit standstills, in which there is no outburst activity \citep{2014JAVSO..42..177S}. Apart from normal outbursts, the so-called superoutbursts, which have an amplitude larger by \( 1^m - 2^m \) and a duration of about two weeks, are observed in SU UMa-type stars \citep{1980A&A....88...66V}. A peculiarity of superoutbursts is the presence of characteristic positive superhumps, i.e., brightness modulations with a period exceeding the orbital one by several percent \citep{1992A&A...263..147H}. The variety of outburst activity in dwarf novae is interpreted in terms of the theory of tidal–thermal accretion disk instability \citep{2001NewAR..45..449L}.

SU UMa-type dwarf novae have an additional internal classification based on the supercycle, i.e., the time between two successive superoutbursts \citep{1996PASP..108...39O, 2001cvs..book.....H}. Classical SU UMa-type systems are characterized by supercycles with a duration of hundreds of days, whereas ER UMa-type stars have supercycles with a duration of \( 20 - 50 \, \text{days} \) \citep{2013arXiv1301.3202K}, and WZ Sge-type stars have very long supercycles reaching tens of years \citep{2007OAP....20..168P}. In addition, there are virtually no normal outbursts in WZ Sge-type dwarf novae \citep{2015PASJ...67..108K}.

In this paper, we present our study of the dwarf nova candidate OGLE-BLG-DN-0064 (hereafter OGLE64) based on archival \textit{Chandra} and \textit{Swift} X-ray data. The optical spectrum of the source OGLE64 was obtained with the 6-m BTA telescope at the Special Astrophysical Observatory of the Russian Academy of Sciences (SAO RAS). We analyzed the outburst activity based on the optical light curves from the Zwicky Transient Facility (ZTF, \citealt{2019PASP..131a8002B}), Optical Gravitational Lensing Experiment (OGLE, \citealt{2003AcA....53..291U}), All-Sky Automated Survey for Supernovae (ASAS-SN, \citealt{2017PASP..129j4502K}), and Asteroid Terrestrial-impact Last Alert System (ATLAS, \citealt{2018PASP..130f4505T}) surveys. In this paper, we discuss the nature of the source OGLE64.

\section{The Source OGLE64}
\begin{figure*}[h!]
    \centering
    \includegraphics[width=0.75\linewidth]{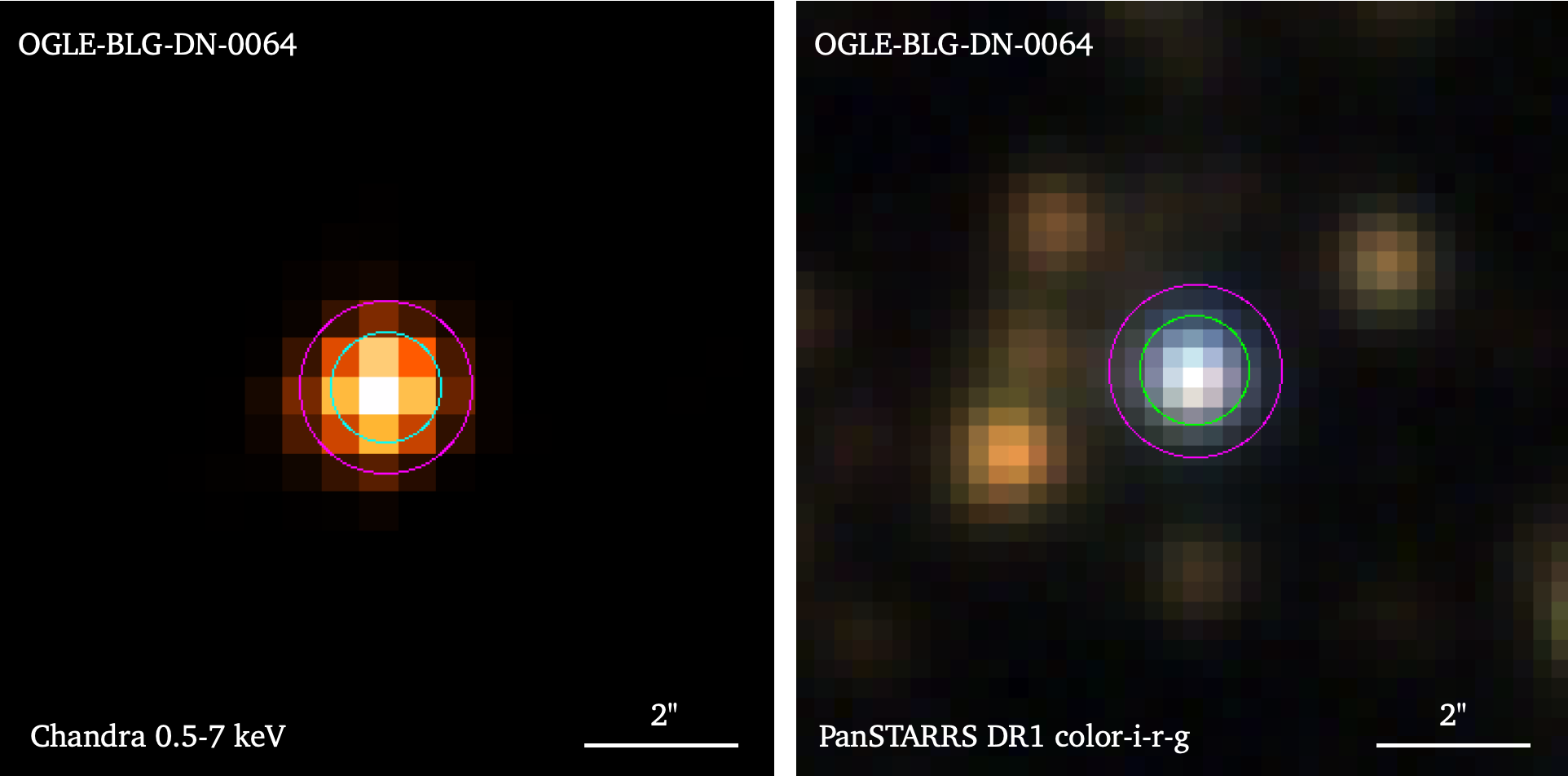}
    \caption{\textit{Left:} The X-ray image of OGLE64 in the 0.5–7 keV energy band from the archival Chandra data (ObsID = 12945). \textit{Right:} The color image of OGLE64 obtained by the superposition of Pan-STARRS images in the gri bands. The inner circle indicates the 95\% position error circle of the X-ray source. The outer circle indicates the region of the point spread function in which 90\% of the signal from the source is expected.}
    \label{fig:X_opt}
\end{figure*}

The source OGLE64 was first noted as a dwarf nova candidate in 2015 owing to its outburst activity recorded by the OGLE optical survey \citep{2015AcA....65..313M}. We found OGLE64 in the \textit{Chandra} Source Catalog 2.0 (CSC2) \citep{2010ApJS..189...37E}  (its name in CSC2 is 2CXO J173917.7–214735). We independently noted it as a potential CV candidate based on its high X-ray-to-optical flux ratio, \( F_X / F_{\text{opt}} \approx 1.5 \), according to the technique proposed by \citet{2024A&A...690A.374G} to search for CVs in CSC2 and the \textit{Gaia} DR3 optical catalog \citep{2023A&A...674A..13E}.

The object OGLE64 coincides only with one \textit{Gaia} DR3 source (source\_id: 4117235609421426560, RA(J2016): \( 17^\text{h}39^\text{m}17.75^\text{s} \), DEC(J2016): \(-21^\circ47'35.56''\)) within a search radius of \( 0.7'' \), which corresponds to the 95\% position error circle of the X-ray source from CSC2. The source OGLE64 is located near the Galactic plane, and, therefore, its absorption and reddening significantly affect the flux from the system. We applied the 3D Bayestar19 map \citep{2019ApJ...887...93G} to estimate the color excess \( E(B-V) = 0.21 \pm 0.01 \, \text{mag} \), which was used in our subsequent analysis. The \textit{Gaia} DR3 parallax of OGLE64 is \( p = (1.15 \pm 0.17) \times 10^{-3} \), corresponding to a distance \( d = 872 \pm 126 \, \text{pc} \). Figure \ref{fig:X_opt} (the right and left panels) shows the X-ray and optical images of OGLE64 constructed from the \textit{Chandra} and Pan-STARRS data \citep{2016arXiv161205560C}.

\section{Analysis of the BTA Optical Spectrum}

\begin{figure*}[h!]
    \centering
    \includegraphics[width=0.75\linewidth]{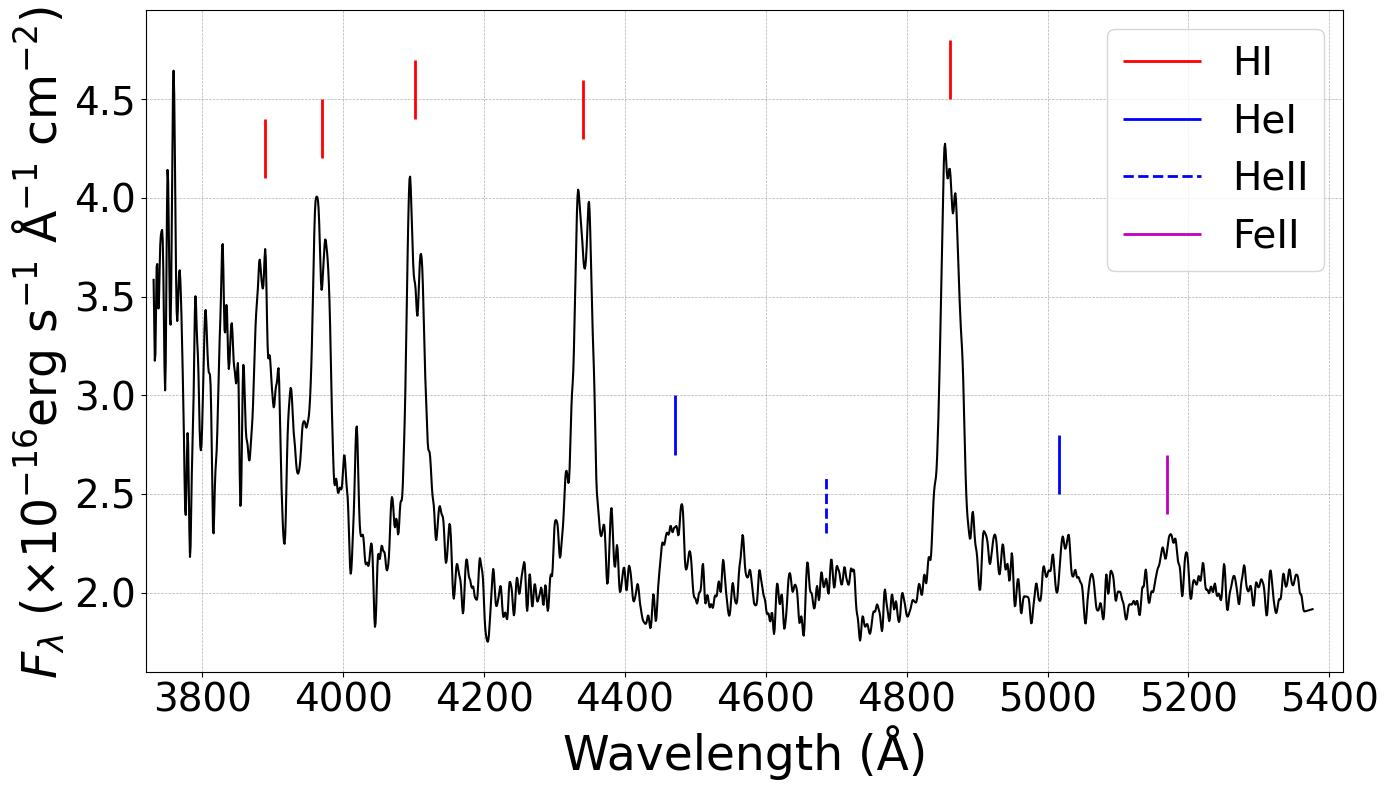}
    \caption{The combined BTA optical spectrum of OGLE64 with the identified emission lines of hydrogen (red bars), neutral helium (blue bars), and ionized iron (violet bars). For better visualization, the optical spectrum was smoothed by the convolution with a Gaussian ($\sigma=1.7$ \AA). A double-peaked structure is observed in some of the emission lines, suggesting the presence of an accretion disk in the system.}
    \label{fig:optspectrum}
\end{figure*}

\begin{figure*}[h]
    \centering
    \includegraphics[width=0.49\linewidth]{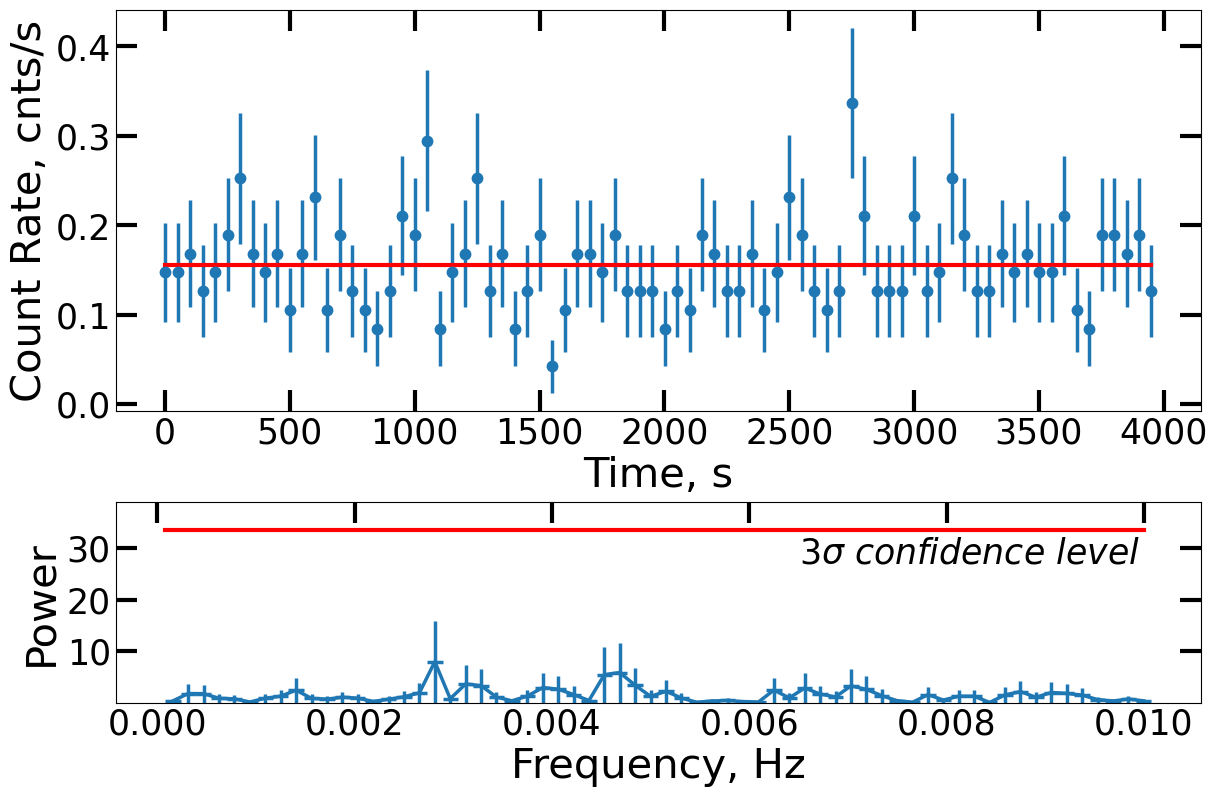}
    \includegraphics[width=0.49\linewidth]{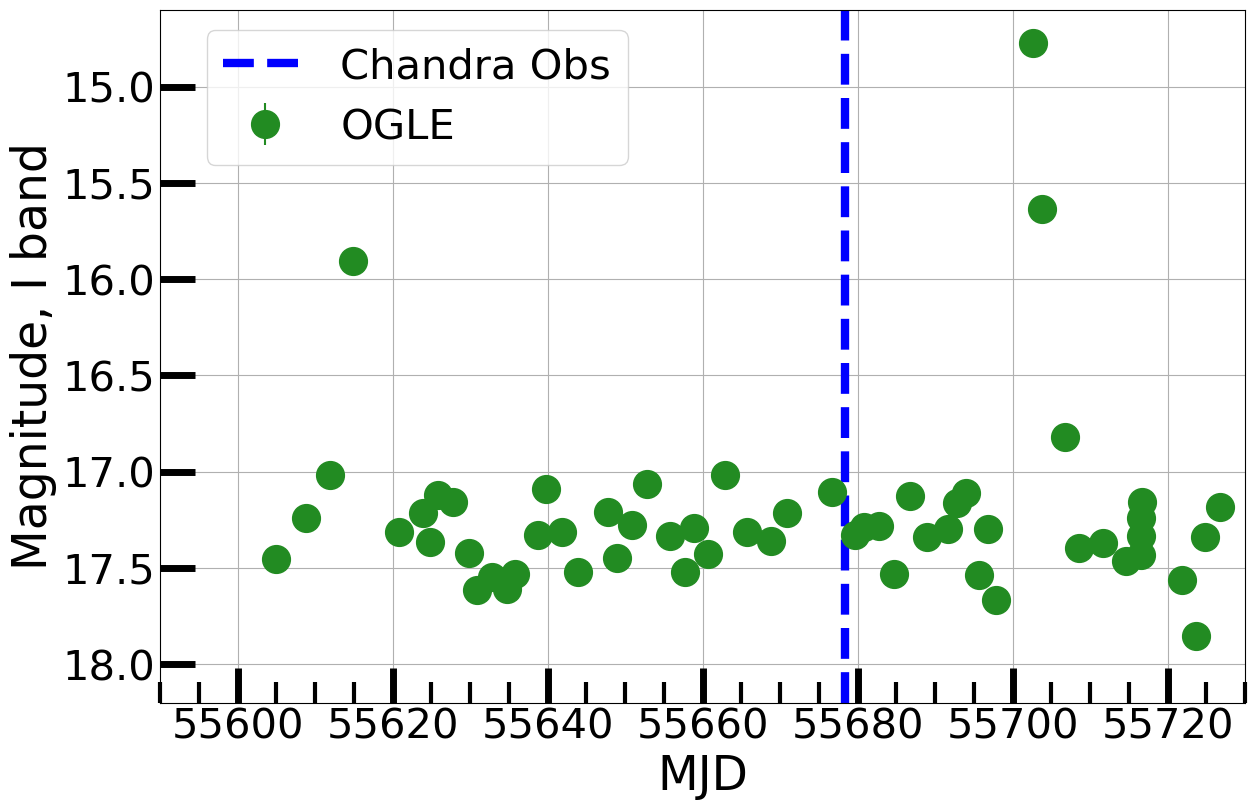}
    \caption{(left) The X-ray light curve of OGLE64 in the 0.5–7 keV energy band with a time resolution of 50 s (upper panel) and the power spectrum (lower panel) constructed from the Chandra data. The red line indicates the mean count rate on the light curve and the power corresponding to the 3$\sigma$ confidence level on the power spectrum. (right) Part of the OGLE optical light curve for OGLE64 covering the Chandra observation. The Chandra observation fell into the quiescent state, approximately 20 days before the outburst.}
    \label{fig:xray_light_curve}
\end{figure*}

Using the 6-m BTA telescope at SAO RAS, we obtained three spectra of OGLE64 with exposure times of 10 min each on the night of June 30, 2024. During the observation, the telescope was equipped with the SCORPIO-1 focal reducer \citep{2005AstL...31..194A} operating in the mode of long-slit spectroscopy. We used the volume phase holographic grating VPHG1200B that covered the spectral range 3600–5400 Å with a resolution \( \Delta\lambda \approx 5.5 \, \text{\AA} \) (the slit width is \( 1.2'' \)). The data reduction was performed using the tools of the IRAF package \citep{1986SPIE..627..733T}, according to the standard technique of working with long-slit spectra. Figure \ref{fig:optspectrum} shows our smoothed combined optical spectrum of OGLE64 with the identified hydrogen, neutral helium, and ionized iron lines, suggesting that OGLE64 is a CV \citep{1980ApJ...235..939W}. A double-peaked emission, typical for the optical spectra of some CVs, is observed in the H\(_\beta\) and H\(_\gamma\) emission lines. The double-peaked emission is a consequence of the Doppler shift caused by the motion of matter in the accretion disk. Table \ref{tab:equivalent_widths} shows the equivalent widths of the emission lines in the combined spectrum calculated by fitting the line profiles with two Gaussians. We calculated the \( 3\sigma \) upper limit for the ratio of the equivalent widths, \( \text{He II}(4686 \, \text{\AA})/\text{H}_\beta \lesssim 0.008 \). Such a low value of this ratio may suggest that OGLE64 is a nonmagnetic CV \citep{1992PhDT.......119S}.

\newcolumntype{C}[1]{>{\centering\arraybackslash}p{#1}}
\begin{table}[t]
\centering
\caption{\small The equivalent widths of the prominent emission lines in the BTA optical spectrum of OGLE64.}
\label{tab:equivalent_widths}
\renewcommand\arraystretch{1.3}
\begin{tabular}{C{0.2\textwidth}|C{0.2\textwidth}}
\hline
\rule{0pt}{12pt}\textbf{Line} & \textbf{$-EW$ (\AA)} \\ \hline
H$_\beta$ & $26.4 \pm 1.4$ \\ 
H$_\gamma$ & $20.2 \pm 1.1$ \\ 
H$_\delta$ & $16.4 \pm 1.0$ \\ 
He~{\sc I} (4471 \AA) & $12.1 \pm 2.4$ \\ 
He~{\sc II} (4686 \AA) & $\lesssim 0.2$ \\ 
Fe~{\sc II} (5169 \AA) & $2.0 \pm 0.2$ \\ \hline
\end{tabular} 

\vspace{5pt}
\noindent\parbox{0.45\textwidth}{
\footnotesize 
The errors in the equivalent widths are given in the \(1\sigma\) confidence interval. For the He II line, we determined the \(3\sigma\) upper limit for the equivalent width.
}
\end{table}

\section{Analysis of the Chandra and Swift X-ray Data}

\begin{figure}[h!]
    \centering
    \includegraphics[width=\linewidth]{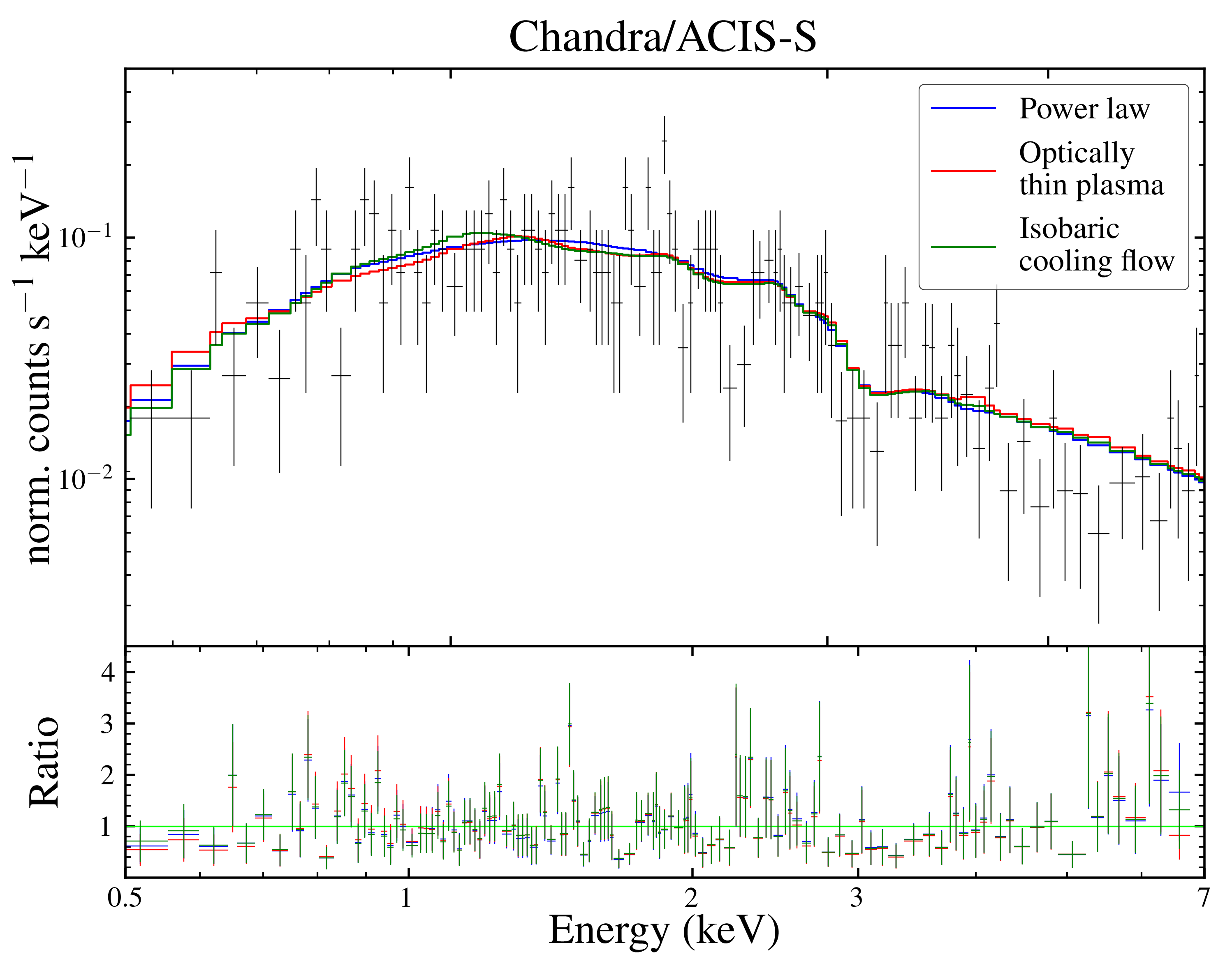}
    \includegraphics[width=\linewidth]{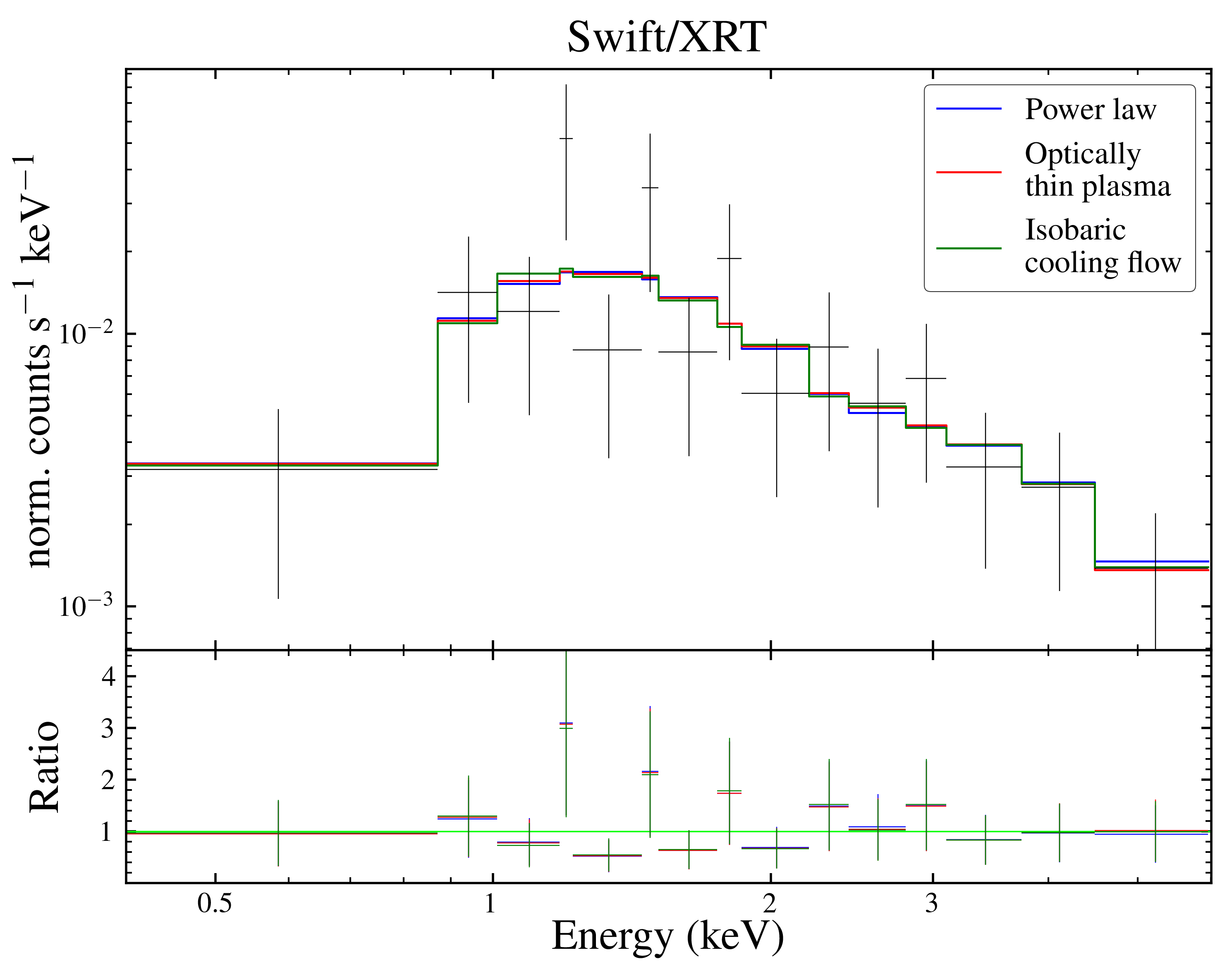}
    \caption{The X-ray spectrum of OGLE64 from the archival Chandra/ACIS-S (\textit{ObsID} = 12945, the upper panel) and Swift/XRT (\textit{ObsID} = 00045781002, the lower panel) data. The different colors represent the power-law (blue line), optically thin plasma (red line), and isobaric cooling flow (green line) models. The panels under each spectrum show the ratio of the observed count rate to the model ones.}
    
    \label{fig:xray_spectra}
\end{figure}

The source OGLE64 fell into the \textit{Chandra} field of view on April 27, 2011 (ObsID = 12945). The primary X-ray data calibration and the extraction of the light curve and spectrum from the event file of the \textit{Chandra} Advanced CCD Imaging Spectrometer S-array (ACIS-S, \citealp{2003SPIE.4851...28G}) observation were performed using the \textit{Chandra} Interactive Analysis of Observations (CIAO) software package \citep{2006SPIE.6270E..1VF}. The X-ray light curve of OGLE64 was extracted from the event file with a time resolution of 50 s (Fig. \ref{fig:xray_light_curve}, left panel). The X-ray telescope onboard the \textit{Swift} space observatory (Swift/XRT, \citealp{2005SSRv..120..165B}) observed the region around OGLE64 on October 30, 2014 and 2015 (ObsID = 00045781002 and 00045781003). For our subsequent analysis, we used only the archival Swift/XRT 2014 data (ObsID = 00045781002), where OGLE64 was detected. We extracted the X-ray spectrum of OGLE64 using the Swift/XRT online service \footnote{\url{https://www.swift.ac.uk/user_objects/}} \citep{2009MNRAS.397.1177E,2020ApJS..247...54E}. The \textit{Chandra}/ACIS-S and Swift/XRT X-ray spectra were binned in such a way that there were at least three counts for each channel. We analyzed the X-ray spectra using the XSPEC software package \citep{1996ASPC..101...17A} in the 0.5–7 keV energy band.

We performed the fast Fourier transform using the XRONOS submodule in FTOOLS (NASA's High Energy Astrophysics Science Archive Research Center, \citep{2014ascl.soft08004N}) in our search for potential periodic signals in the \textit{Chandra} X-ray light curve. To estimate the significance of the peaks in the constructed periodogram, we generated 2000 random light curves with the mean count rate of OGLE64 by assuming that the X-ray flux from OGLE64 was constant in time. Subsequently, we reapplied the fast Fourier transform to each of the randomly generated light curves, determined the powers of the maximal peaks, and constructed their distribution. Based on this distribution, we calculated the power of the peak corresponding to the \(3\sigma\) confidence level. As a result of our analysis, we found no significant periodicity in the \textit{Chandra} observed X-ray light curve (Fig. \ref{fig:xray_light_curve}, left panel). The absence of rotational flux modulation in the X-ray light curve of OGLE64 may be indicative of a nonmagnetic nature of the system.

The \textit{Chandra}/ACIS-S and Swift/XRT X-ray spectra were fitted by three different models: a power law (\texttt{powerlaw} in XSPEC), the model of an optically thin plasma (\texttt{mekal} in XSPEC, \citealt{1986A&AS...65..511M, 1995ApJ...438L.115L}), and the model of an isobaric cooling flow (\texttt{mkcflow} in XSPEC, \citealt{1988ASIC..229...53M}). The component responsible for the interstellar absorption according to the Tübingen–Boulder law with the chemical composition derived by \citet{2000ApJ...542..914W} was added to the models. The C-statistic \citep{1979ApJ...228..939C} was used to fit the spectra. Figure \ref{fig:xray_spectra} shows the \textit{Chandra}/ACIS-S and Swift/XRT X-ray spectra of OGLE64 and the different models used to fit these spectra. Table \ref{tab:xray_fit_results} presents the results of fitting the \textit{Chandra}/ACIS-S and Swift/XRT X-ray spectra of OGLE64 by the different models. For each model, we provide the ratio of the C-statistic to the number of degrees of freedom (C-stat/dof) obtained as a result of our fitting. We used the \texttt{error} command in XSPEC to calculate the errors of the parameters in the \(1\sigma\) confidence interval. We calculated the absorption-corrected X-ray fluxes and luminosities in the 0.5–7 keV energy band using the power-law model.

The results of fitting the \textit{Chandra}/ACIS-S and Swift/XRT X-ray spectra by the different models agree between themselves, and no statistically significant deviations at the \(3\sigma\) confidence level were found (see Table \ref{tab:xray_fit_results}). The signal-to-noise ratio of the \textit{Chandra}/ACIS-S X-ray spectrum is larger than that of the Swift/XRT spectrum. The \textit{Chandra}/ACIS-S spectrum is better fitted by the power-law model with a hydrogen column density \(N_H = 3.28^{+0.65}_{-0.63} \times 10^{21} \, \text{cm}^{-2}\) and a photon index \(\Gamma = 1.85 \pm 0.13\) (C-stat/dof = 118.82/124). The fit by the optically thin plasma model is characterized by a hydrogen column density \(N_H = 2.07^{+0.45}_{-0.43} \times 10^{21} \, \text{cm}^{-2}\) and a plasma temperature \(kT = 6.44^{+1.82}_{-1.22} \, \text{keV}\) (C-stat/dof = 122.81/124). Fitting the \textit{Chandra} spectrum by the isobaric cooling flow model gives the following parameters: \(N_H = 2.79^{+0.54}_{-0.70} \times 10^{21} \, \text{cm}^{-2}\), \(kT_{\text{max}} = 18.93^{+13.93}_{-5.84} \, \text{keV}\), and \(\dot{M}_{\text{acc}} = 5.69^{+2.05}_{-1.89} \times 10^{-11} \, M_{\odot} \, \text{yr}^{-1}\) (C-stat/dof = 119.17/122). The X-ray luminosity (\(L_X \approx 1.6 \times 10^{32} \, \text{erg s}^{-1}\)), the photon index, the optically thin plasma temperature, and the accretion rate obtained by fitting the X-ray spectra of OGLE64 are typical for nonmagnetic CVs in quiescence \citealp{2021AstL...47..587G}.

Figure \ref{fig:xray_light_curve} (right panel) shows part of the OGLE optical light curve for OGLE64 with the marked start time of the \textit{Chandra}/ACIS-S observation that fell into the quiescent state of OGLE64. The Swift/XRT observation time fell into the interval for which there are no optical data. However, based on our comparison of the \textit{Chandra}/ACIS-S and Swift/XRT X-ray spectra, we can assume that OGLE64 was also in quiescence.

\begin{table*}[h]
    \caption{ The results of fitting the Chandra/ACIS-S and Swift/XRT X-ray spectra of OGLE64 in the 0.5–7 keV energy band}
    \centering
    \fontsize{9}{10}\selectfont
    \renewcommand\arraystretch{1.6}
    \setlength\tabcolsep{0pt}
    \begin{tabular}{C{0.4\textwidth}|C{0.3\textwidth}|C{0.3\textwidth}} \hline
          & Chandra/ACIS-S & Swift/XRT \\ 
         Date & April 27, 2011 & October 30, 2014 \\ 
         ObsID & 12945 & 00045781002 \\ \hline
         \multicolumn{3}{c}{Power law} \\ \hline 
         $N_H$, $\times10^{21}$ cm$^{-2}$ & $3.28^{+0.65}_{-0.63}$ & $5.41^{+4.09}_{-2.95}$ \\
         $\Gamma$& $1.85\pm 0.13$ & $1.66^{+0.54}_{-0.49}$ \\
         C-stat/dof & 118.82/124 & 8.84/12 \\ 
         $F_{0.5-7}$, $\times 10^{-12}$ erg s$^{-1}$ cm$^{-2}$ & $1.73 \pm 0.10$ & $2.74 \pm 0.72$ \\ 
         $L_{0.5-7}$, $\times 10^{32}$ erg s$^{-1}$ & $1.58 \pm 0.47$ & $2.49 \pm 0.97$ \\ \hline
         \multicolumn{3}{c}{Optically thin plasma model} \\ \hline 
         $N_H$, $\times10^{21}$ cm$^{-2}$ & $2.07^{+0.45}_{-0.43}$ & $4.72^{+3.17}_{-2.33}$ \\
         $kT$, keV & $6.44^{+1.82}_{-1.22}$ & $\geq 3.66$ \\
         C-stat/dof & 122.81/124 & 8.75/12 \\ \hline 
         \multicolumn{3}{c}{Isobaric cooling model} \\ \hline 
         $N_H$, $\times10^{21}$ cm$^{-2}$ & $2.79^{+0.54}_{-0.70}$ & $4.66^{+3.15}_{-2.28}$ \\
         $kT_{max}$, keV & $18.93^{+13.93}_{-5.84}$ & $\geq 3.76$ \\
         $\dot{M}_{acc}$, $\times 10^{-11}M_\odot$ yr$^{-1}$ & $5.69^{+2.05}_{-1.89}$ & $\leq 264$ \\
         C-stat/dof & 119.17/122 & 7.61/10 \\ \hline
         
    \end{tabular}
        \vspace{1pt}
    \vspace{3pt}
    \noindent\parbox{\textwidth}{
        \footnotesize The X-ray fluxes and luminosities were corrected for the absorption. The errors are given in the 1$\sigma$. confidence interval.
    }
    \label{tab:xray_fit_results}
\end{table*}

\section{Analysis of the Optical Light Curves}

\begin{figure*}[h!]
    \centering
    \includegraphics[width=1.0\linewidth]{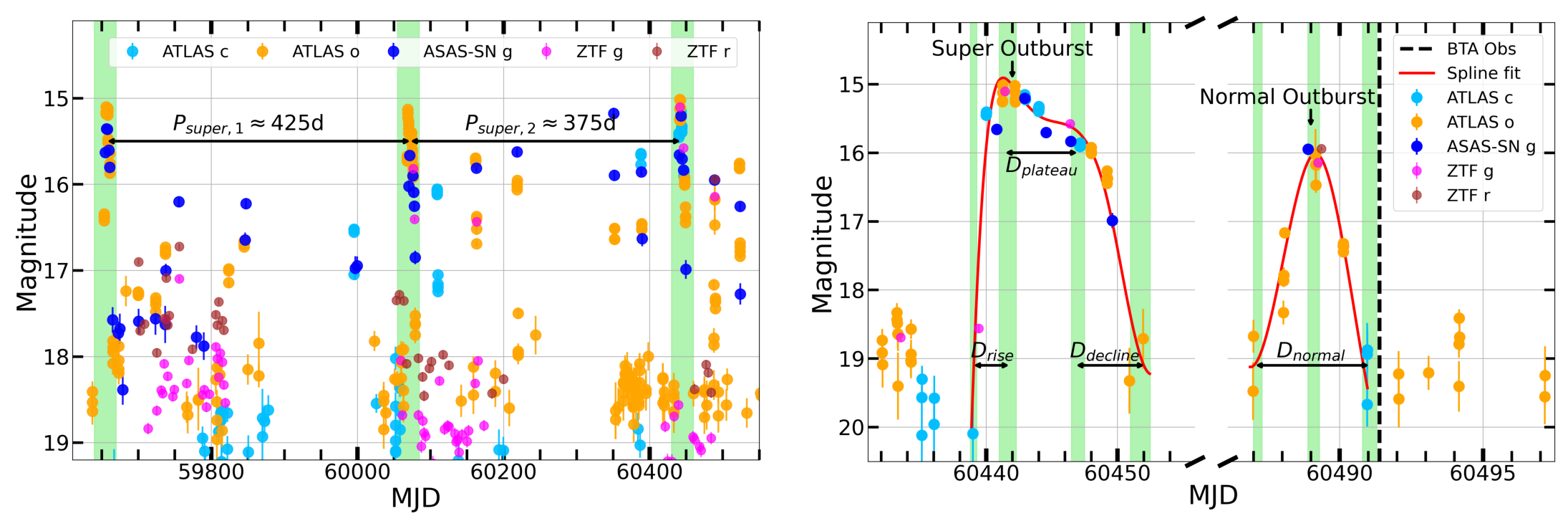}
    \caption{The ATLAS, ASAS-SN, and ZTF optical light curves of OGLE64 in the $g,r,o,c$ bands. (left) The optical light curve showing three successive superoutbursts. The green bands indicate the times of the superoutbursts with characteristic supercycles $P_{super, 1} \approx 425$ and $P_{super, 2} \approx 375$. (right) Part of the optical light curve showing two successive outbursts with different morphologies. The period of the quiescent state between the outbursts was excluded. For better visualization of the differences in morphology, the outbursts were smoothed using a spline function (red line). The green bands mark the boundaries of the various outburst phases. The black dashed line indicates the BTA observation time.}
    \label{fig:Outbursts}
\end{figure*}
The source OGLE64 was observed in several op- tical surveys, including ZTF, OGLE, ASAS-SN, and ATLAS. The left and right panels of Fig. \ref{fig:Outbursts} present the ATLAS, ASAS-SN, and ZTF optical light curves of OGLE64 in the \(g\), \(r\), \(o\) (560–820 nm), and \(c\) (420–650 nm) bands.

The optical light curves of OGLE64 exhibit an outburst activity that manifests itself as an increase in the brightness of the system by \(\sim 3^m - 5^m\) in different bands. Superoutbursts with larger amplitudes and durations are also observed in the light curves. The left panel of Fig. \ref{fig:Outbursts} shows three recent optical superoutbursts. The intervals between two successive superoutbursts are \(P_{\text{super}, 1} \approx 425 \, \text{days}\) and \(P_{\text{super}, 2} \approx 375 \, \text{days}\), which allows the average supercycle of the system to be determined as \(P_{\text{super}} \approx 400 \, \text{days}\). Based on our estimate of the supercycle, one might expect that the next superoutburst will occur in the middle of 2025. The presence of such superoutbursts suggests that the source OGLE64 belongs to the subclass of SU UMa-type dwarf novae \citep{2001cvs..book.....H}.

The right panel of Fig. \ref{fig:Outbursts} presents part of the optical light curve in which two successive outbursts with different morphologies and a difference in time of about 35 days are observed. The first outburst (superoutburst) is characterized by a longer duration and an extended plateau phase observed after reaching the brightness peak. In contrast to it, the second (normal) outburst has a triangular shape, a shorter duration, and a brightness lower approximately by \(\sim 1^m\). The symmetric triangular shape of normal outbursts typical for SU UMa-type dwarf novae suggests that the outburst begins in the inner regions of the accretion disk and propagates toward its outer boundaries \citep{2010ApJ...725.1393C}.

We analyzed the various parameters related to the outburst amplitude and time. For the superoutbursts, we determined the rise time, \(D_{\text{rise}} \approx 3.0 \, \text{days}\), the duration of the plateau phase, \(D_{\text{plateau}} \approx 5.8 \, \text{days}\), and the decline time, \(D_{\text{decline}} \approx 5.5 \, \text{days}\). The magnitudes at the superoutburst peak and in quiescence are \(m_{\text{max}, \text{super}} \approx 15^m\) and \(m_{\text{min}} \approx 19.5^m\), respectively. For the normal outbursts, we determined the total outburst time, \(D_{\text{normal}} \approx 4.1 \, \text{days}\), and the magnitude at the peak, \(m_{\text{max}, \text{normal}} \approx 16^m\).

We searched for a periodicity in the light curves to determine the orbital period of the system. To remove the data points corresponding to the outbursts, we applied asymmetric sigma-clipping\footnote{\url{https://docs.astropy.org/en/stable/api/astropy.stats.sigma_clip.html}} to the light curves. For each optical light curve, we constructed the Lomb–Scargle periodogram \citep{1976Ap&SS..39..447L, 1982ApJ...263..835S}. The significance of the peaks in the periodograms was determined by the bootstrap method \citep{2018ApJS..236...16V}. As a result of our analysis, we found no significant (\(3\sigma\)) periodicity in any of the ZTF, OGLE, ASAS-SN, and ATLAS light curves used. The absence of significant peaks in the periodograms may be related to the small data volume. In addition, in the available data for the superoutbursts, we observe no positive superhumps typical for SU UMa-type dwarf novae, which is most likely related to the data sampling.

\section{Discussion}

\subsection{An Estimate of the WD Mass}

To estimate the WD mass, we used the following relation between the shock temperature and the WD mass\citep{2002apa..book.....F}:

\[
kT_{\text{max}} = \alpha \times \frac{3}{16} \times \frac{GM_{\text{WD}} m_{\text{H}} \mu}{R_{\text{WD}}}.
\]

Here, \(\alpha = 0.646\) is an empirical calibration constant \citep{2018ApJ...853..182Y, 2022RNAAS...6...65M}, \(kT_{\text{max}}\) is the shock temperature (we used the value obtained when fitting the X-ray spectrum by the isobaric cooling model), \(m_{\text{H}}\) is the mass of the hydrogen atom, \(\mu\) is the mean molecular weight (\(\mu = 0.615\)), \(G\) is the gravitational constant, and \(M_{\text{WD}}\) and \(R_{\text{WD}}\) are the WD mass and radius, respectively. Using the mass–radius relation for WDs \citep{1972ApJ...175..417N}, we obtained the following estimate for the WD mass:
\(
M_{\text{WD}} \approx 1.00^{+0.21}_{-0.16} \, M_{\odot}.
\)

\subsection{Estimates of the Orbital Period and the Mass Ratio in the System}

We found no significant variability in the ATLAS, OGLE, ASAS-SN, and ZTF optical light curves of OGLE64 that could be associated with the change in the visibility conditions for the emitting regions at different orbital phases. In spite of this, we used several indirect methods to estimate the orbital period of the system and the donor parameters. Table \ref{tab:orbital_period_estimates} gives various estimates of the orbital period and the mass ratio in the system.

\begin{table}[h]
\centering
\caption{Estimates of the orbital period and the mass ratio in OGLE64 by different methods.}
\label{tab:orbital_period_estimates}
\renewcommand{\arraystretch}{1.4} 
\small
\begin{tabular}{p{0.2\textwidth}|C{0.1\textwidth}|C{0.11\textwidth}}
\hline
\textbf{Method} & \textbf{\(P_{\text{orb}}\) (h)} & \textbf{\(q\)} \\
\hline
1. The peak value of \(M_V\) & \(\sim 1.7 - 3.5\) & \(\sim 0.03 - 0.29\) \\
\hline
2. The normal outburst  & \multirow{2}{*}{$\sim 0.6 - 3.3$} & \multirow{2}{*}{-} \\ 
duration & \\ \hline
3. The object position & \multirow{2}{*}{$\sim 2.6 - 3.7$} & \multirow{2}{*}{-}\\ 
on the HR diagram & \\ \hline
4. The semi-empirical & \multirow{2}{*}{$\lesssim 4$} & \multirow{2}{*}{$\lesssim 0.4$} \\
donor sequence & \\
\hline
\end{tabular}

\vspace{0.2cm}
\footnotesize
\noindent\parbox{0.45\textwidth}{The estimates of the orbital period and mass ratio are based on indirect methods described in the text. The mass ratio \(q\) is defined as \(q = M_2 / M_{\text{WD}}\), where \(M_2\) is the donor mass and \(M_{\text{WD}}\) is the white dwarf mass.}
\end{table}

\subsubsection{Method 1: The Peak Value of \(M_V\)}

For dwarf novae, there is a relationship between the peak absolute magnitude during the outburst and the orbital period, \(M_{V,\text{max}}(P_{\text{orb}})\) \citep{2011MNRAS.411.2695P}. We calculated the mean magnitude at the superoutburst peak, \((m_g)_{\text{max}} = 15.11 \pm 0.05\), and converted the magnitude from the \(g\) band to the \(V\) one \citep{2005AJ....130..873J} based on the (BP-RP) color index from the \textit{Gaia} data\footnote{\url{https://gea.esac.esa.int/archive/documentation/CDR3/Data_processing/chap_cu5pho/cu5pho_sec_photSystem/cu5pho_ssec_photRelations.html}}. Assuming the Cardelli extinction law (\(R_V = 3.1\), \cite{}) and using \(E(B-V)\) from the Bayestar19 map \cite{}, we calculated the correction for the interstellar extinction, \(A_V = 0.66 \pm 0.04\). As a result, we obtained the absolute magnitude at the peak, \((M_V)_{\text{max}} = 4.42 \pm 0.07\). Using Eq. (4) from \citet{2011MNRAS.411.2695P}, we estimated the orbital period of OGLE64 to be \(P_{\text{orb}} \sim 1.7 - 3.5 \, \text{h}\).

\subsubsection{Method 2: The Normal Outburst Duration}

We used the relationship between the duration of a normal outburst and the orbital period, \(D_{\text{normal}}(P_{\text{orb}})\) \citep{2000NewAR..44..171S}, where for OGLE64 the normal outburst duration is \(D_{\text{normal}} \approx 4.1 \, \text{days}\). For the parameter \(D_{\text{normal}}\), we adopted a typical error in the outburst duration of about 0.4 day due to the optical data sampling. Using Eq. (19) from \citet{2016MNRAS.460.2526O}, we estimated the orbital period to be \(P_{\text{orb}} \sim 0.6 - 3.3 \, \text{h}\). From the relationship between the outburst duration and the component mass ratio in the system \(D_{\text{normal}}(q)\), we estimated the mass ratio for OGLE64 to be \(q \sim 0.03 - 0.29\) \citep{2016MNRAS.460.2526O}.

\begin{figure}[h!]
    \centering
    \includegraphics[width=\linewidth]{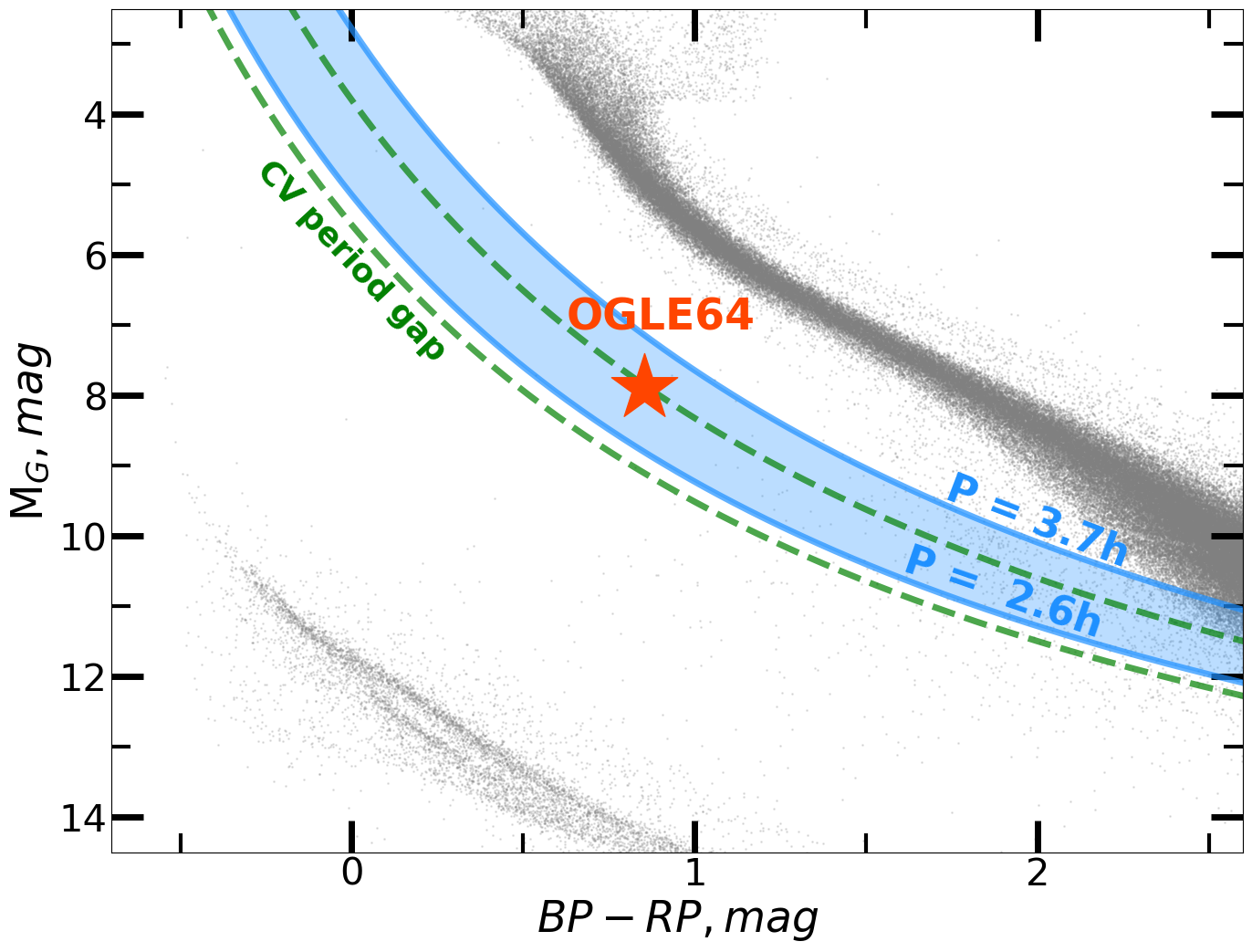}
    \caption{The HR diagram constructed on the basis of stars from the Gaia catalog within 100 pc with statistically significant measured parallaxes (S/N > 100) \citep{2021A&A...649A...6G}.  The position of OGLE64 is marked by the red star. The blue color
    indicates the characteristic position region of CVs with orbital periods $P_{orb} \sim 2.6 - 3.7$ h from \citet{2022ApJ...938...46A}. The green lines indicate the position regions of CVs with orbital periods $P_{1,2} = 2.45, 3.18$ h corresponding to the upper and lower boundaries of the CV period gap from \citet{2024A&A...682L...7S}.}
    \label{fig:hr_diagram}
\end{figure}

\subsubsection{Method 3: The Object Position on the HR Diagram}

We used the relationship between the source position on the Hertzsprung–Russell (HR) diagram and the orbital period \citep{2020MNRAS.492L..40A, 2022ApJ...938...46A}. We assumed that the magnitudes of OGLE64 \(m_G = 18.10\), \(m_{BP} = 18.43\), and \(m_{RP} = 17.31\) were obtained by \textit{Gaia} in quiescence. Using the conversion factors \(R(a) = A(a)/E(B-V)\) from \citet{2023ApJS..264...14Z}, we calculated the extinction in the \textit{Gaia} bands: \(A_G = 0.50 \pm 0.03\), \(A_{BP} = 0.64 \pm 0.04\), and \(A_{RP} = 0.37 \pm 0.02\). As a result, we calculated the absolute magnitude, \(M_G = 7.90 \pm 0.03\), and the color index, \(BP - RP = 0.85 \pm 0.10\). We used Eq. (1) from \citet{2022ApJ...938...46A} to estimate the orbital period of OGLE64, \(P_{\text{orb}} \sim 2.6 - 3.7 \, \text{h}\). Figure \ref{fig:hr_diagram} shows the position of OGLE64 on the HR diagram and the characteristic position region of CVs with orbital periods \(P_{\text{orb}} \sim 2.6 - 3.7 \, \text{h}\) from \citet{2022ApJ...938...46A}. 

\subsubsection{Method 4: The Semi-Empirical Donor Sequence}

We compared the observed near-infrared magnitudes of OGLE64 with the values obtained from the semi-empirical donor sequence \citep{2006MNRAS.373..484K, 2011ApJS..194...28K}. Two observations of OGLE64 in quiescence are available in the Vista Variables in the Via Lactea DR4.2 catalog \citep{2010NewA...15..433M,2023yCat.2376....0M}. The mean magnitudes in the catalog corrected for the interstellar extinction are \(m_J = 16.54 \pm 0.04\), \(m_H = 16.24 \pm 0.05\), and \(m_{Ks} = 16.02 \pm 0.06\). We converted these magnitudes to the photometric system of the California Institute of Technology (CIT)\footnote{\url{https://www.astro.caltech.edu/~jmc/2mass/v3/transformations/}}, and calculated the absolute magnitudes \(M_J = 6.82 \pm 0.12\), \(M_H = 6.59 \pm 0.10\), and \(M_K = 6.33 \pm 0.06\). We compared these magnitudes with the semi-empirical donor sequence and estimated the upper limits for the following parameters of OGLE64: the orbital period \(P_{\text{orb}} \leq 4 \, \text{h}\), the donor mass \(M_2 \lesssim 0.3 \, M_{\odot}\), the donor radius \(R_2 \lesssim 0.4 \, R_{\odot}\), and the effective temperature of the donor \(T_2 \lesssim 3500 \, \text{K}\). We used the previously estimated WD mass and set the upper limit for the mass ratio \(q \lesssim 0.4\). It is important to note that we assumed that all of the infrared radiation is generated only by the donor star, and, therefore, our estimates of the parameters for the donor star are the limiting ones.

\subsection{The Evolutionary Status of OGLE64}

\begin{figure}[h!]
    \centering
    \includegraphics[width=\linewidth]{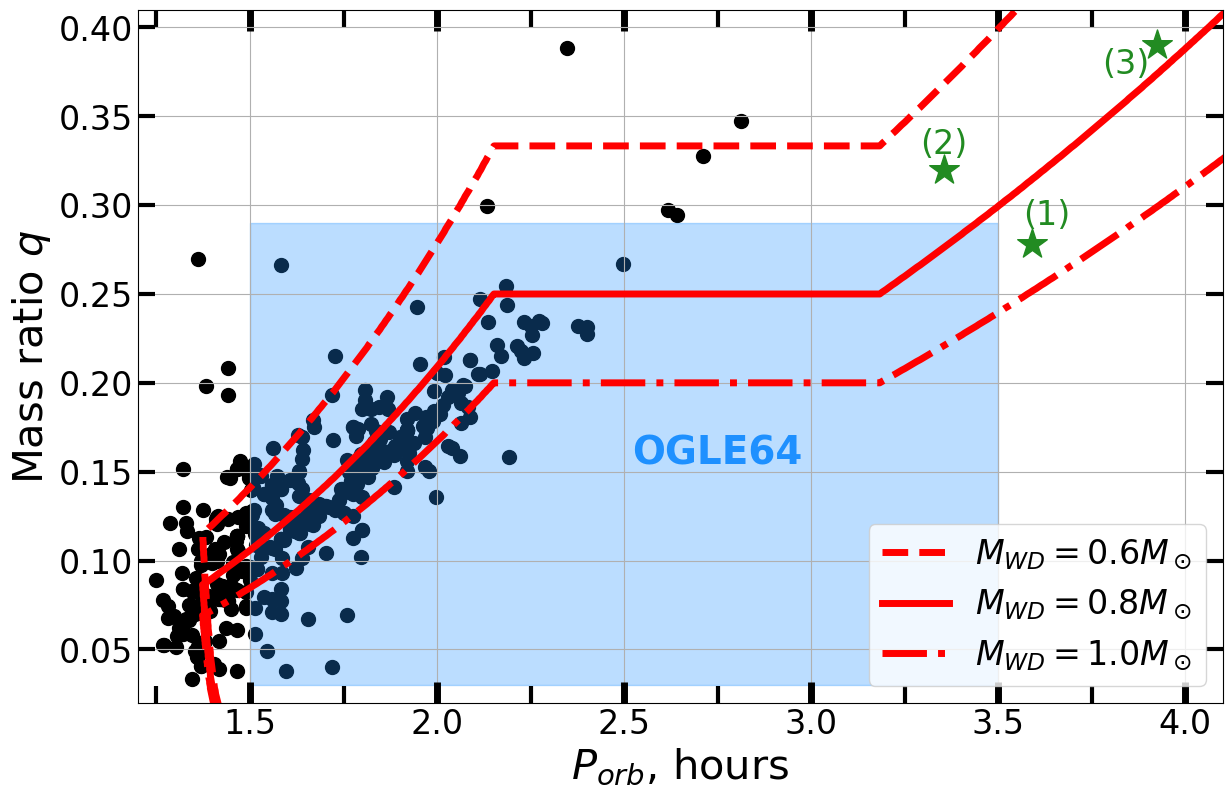}
    \caption{The mass ratio–orbital period diagram for known SU UMa-type dwarf novae from \citet{2016MNRAS.460.2526O}. The blue color marks the position region of OGLE64. We constructed the red lines based on the semi-empirical donor sequence \citep{2011ApJS..194...28K} with several WD masses. The green stars mark some of the known dwarf novae: (1) ASASSN-18aan \citep{2021PASJ...73.1209W}; (2) BO Cet \citep{2021PASJ...73.1280K}; (3) SDSS J094002.56+274942.0 \citep{2023arXiv230413311K}.} 
    \label{fig:mass_ratio_orbital_period}
\end{figure}

The presence of superoutbursts in OGLE64 confirms the nature of the source as an SU UMa-type dwarf nova (see the Section on the analysis of the optical light curves). The most probable range for the orbital period is \(P_{\text{orb}} \sim 1.5 - 3.5 \, \text{h}\), while the range for the mass ratio in the system is \(q \sim 0.03 - 0.29\).

In Fig. \ref{fig:mass_ratio_orbital_period}, the mass ratio is plotted against the orbital period for known SU UMa-type dwarf novae. The red lines correspond to the semi-empirical donor sequence \citep{2011ApJS..194...28K} for which we calculated the mass ratio by assuming different WD masses (\(M_{\text{WD}} = 0.6, 0.8, 1.0 \, M_{\odot}\)) in the CV. As can be seen from Fig. \ref{fig:mass_ratio_orbital_period}, the general population of SU UMa-type dwarf novae has an orbital period below the period gap and a mass ratio \(q \lesssim 0.25\). However, we know about ten systems located in the period gap and exceeding the limiting value of \(q\). The reasons for their stay in this state still remain unknown \citep{1984A&A...132..187S, 2014PASJ...66..111P, 2018AJ....155..232L}. In addition, SU UMa-type dwarf novae with a high mass ratio located above the period gap have been discovered recently, which distinguishes them from the general population still more (see Fig. \ref{fig:mass_ratio_orbital_period}): ASASSN-18aan  (\(P_{\text{orb}} \approx 3.59 \, \text{h}\) and \(q \approx 0.28\), \cite{2021PASJ...73.1209W}), BO Cet ($P_{orb} \approx 3.36$ h and $q \approx 0.32$, \citealp{2021PASJ...73.1280K}), and SDSS J094002.56+274942.0 ($P_{orb} \approx 3.92$ h and $q \approx 0.39$, \citealp{2023arXiv230413311K}). Our estimates of the orbital period of OGLE64 and the mass ratio in the system admit that OGLE64 may be in the CV period gap or closer to its upper boundary, while having a higher mass ratio than the general population of SU UMa-type dwarf novae (see Figs. \ref{fig:hr_diagram} and \ref{fig:mass_ratio_orbital_period}).

\section{Conclusions}

OGLE64 was noted as a potential dwarf nova candidate based on its outburst activity revealed by the OGLE optical survey. We found the source OGLE64 in CSC2 and independently noted it as a potential candidate for accreting WDs based on its high X-ray-to-optical flux ratio \(F_X/F_{\text{opt}} \approx 1.5\). We obtained the following results:

\begin{itemize}
\item[---] OGLE64 shows an X-ray luminosity \(L_X \approx 1.6 \times 10^{32} \, \text{erg s}^{-1}\) in the 0.5-7 keV energy band. \textit{Chandra} and \textit{Swift} X-ray spectra of OGLE64 are better fitted by the models of a power law with a photon index \(\Gamma \approx 1.9\) and an optically thin plasma with a temperature \(kT \approx 6.4 \, \text{keV}\). Such spectral parameters are typical for nonmagnetic CVs. We found no significant differences in the spectral parameters and X-ray luminosities of OGLE64 between the \textit{Chandra} and \textit{Swift} observations. The isobaric cooling flow model allows us to estimate the accretion rate in the system as \(\dot{M}_{\text{acc}} \approx 5.4 \times 10^{-11} M_{\odot} \, \text{yr}^{-1}\).

\item[---] The optical spectrum of OGLE64 obtained using the 6-m BTA telescope at the Special Astrophysical Observatory of the Russian Academy of Sciences exhibits emission lines of hydrogen and neutral helium. A double-peaked emission is observed in some of the hydrogen lines, suggesting the presence of an accretion disk in the system. The ratio of the equivalent widths \(\text{He II}(4686 \text{Å})/\text{H}_\beta \lesssim 0.008\) may suggest that OGLE64 is a nonmagnetic CV.

\item[---] The ATLAS, ASAS-SN, OGLE, and ZTF optical light curves exhibit an outburst activity that manifests itself as an increase in the brightness of OGLE64 by \(\sim 3^m - 5^m\). We detected superoutbursts with a characteristic supercycle \(P_{\text{super}} \approx 400 \, \text{days}\). We found no significant periodicity in the X-ray and optical light curves, which may suggest that the system is more likely a non-eclipsing one. Our estimates of the system's orbital period by indirect methods indicate that the period of OGLE64 probably lies in the range \(P_{\text{orb}} \sim 1.5-3.5 \, \text{h}\).

\item[---] The properties of the X-ray and optical emissions lead us to conclude that the source OGLE64 is a nonmagnetic CV. The outburst activity of the system suggests that OGLE64 is an SU UMa-type dwarf nova.
\end{itemize}

\acknowledgements

\section{Acknowledgements}

In this study we used the data retrieved from the Chandra data archive and the Chandra source catalog as well as the software provided by the Chandra X-ray Center (CXC) in the CIAO and Sherpa applications. We thank the members of the OGLE team for their contribution to the OGLE photometric data acquisition. We used data from the Gaia mission of the European Space Agency (ESA) (https://www.cosmos.esa.int/gaia) processed by the Gaia Data Processing and Analysis Consor- tium (DPAC, https://www.cosmos.esa.int/web/ gaia/dpac/consortium). DPAC is funded by the national institutions, in particular, the institutions participating in the Gaia multilateral agreement.

\section{Funding}

The observations at the SAO RAS telescopes are supported by the Ministry of Science and Higher Education of the Russian Federation. The instru- mentation is upgraded within the “Science and Uni- versities” National Project. This work was supported by the Kazan Federal University.

\section{Conflicts of interest}

The authors of this work declare that they have no
conflicts of interest.

\bibliographystyle{astl}
\bibliography{refs2}

\end{document}